\documentclass{article}
\usepackage{textcomp}
\usepackage{siunitx}
\usepackage[utf8]{inputenc}
\usepackage{graphicx} 
\usepackage{textgreek} 
\usepackage{amsmath} 

\usepackage{gensymb} 
\usepackage{float} 
\usepackage{booktabs}
\usepackage{threeparttable}

\usepackage{makecell}
\usepackage[
backend=biber,
style=apa
]{biblatex}

\DeclareLanguageMapping{english}{english-apa}

\usepackage{hyperref}
\usepackage{authblk}

\usepackage{appendix}

\title{Analyzing the daily flows: Exploring shared micro-mobility factors in Venice}

\author[1]{Vanshika Keshwani}
\author[1]{Stefano Mazzuco}

\affil[1]{Department of Statistical Sciences, University of Padua, Italy}

\date{}

\begin{document}
	
	\begin{titlepage}
		
		\begin{center}
			{\Large \textbf{Analyzing the daily flows: Exploring shared micro-mobility factors in Venice}}
		\end{center}
		
		\vspace{1cm}
		
		\noindent
		\textbf{Author Names:}
		
		\vspace{0.2cm}
		
		\noindent
		Vanshika Keshwani$^{1}$, Stefano Mazzuco$^{1}$
		
		\vspace{0.5cm}
		
		\noindent
		\textbf{Affiliation:}
		
		\vspace{0.2cm}
		
		\noindent
		$^{1}$ Department of Statistical Sciences, University of Padua, Italy
		
		\vspace{0.5cm}
		
		\noindent
		\textbf{Corresponding author's address}
		
		\vspace{0.2cm}
		
		\noindent
		Vanshika Keshwani\\
		Department of Statistical Sciences\\
		University of Padua\\
		35121 Padua, Italy\\
		Email: vanshika.keshwani@studenti.unipd.it

	\end{titlepage}
	
	\maketitle
	
	
	
	\begin{abstract}
Shared micro-mobility has emerged as a key component of a sustainable urban transportation system, however, limited research exists on how environmental factors influence the mobility demand between specific origin-destination (OD) locations. This work extends research on the demand-side perspective to explore how temporal and environmental conditions shape daily shared micro-mobility flows in Venice. The study analyses repeated variation across 158,401 OD-day observations for two years in 50 spatial zones. Daily temperature, rainfall, and PM$_{10}$ concentrations are linked to each OD-day observation while accounting for vehicle-pass composition and temporal patterns. Here, the unit of analysis is the connection between OD pairs. A generalised additive mixed model (GAMM) is used to represent the non-linearity of environmental relationships across seasons, providing a flexible framework for understanding how climatic conditions influence sustainable mobility behaviour. The results show a significant nonlinear association between temperature and mobility demand across seasons. High rainfall is associated with reduced demand, with larger reductions under moderate and heavy rainfall than on dry days. The relationship between PM$_{10}$ and shared mobility use was season-dependent, creating an avoidance-versus-adoption mechanism rather than a monotonic association. After adjustment for environmental and temporal factors, a recurring increase in demand within the Lido Islands during August and September remained evident, highlighting a location-specific mobility pattern across two years. The study highlights the importance of environmental sensitivity in shared micro-mobility research. This work illustrates that the adoption of shared bikes and electric bikes depends not only on service availability but also on usage patterns, which are affected by external conditions.

\end{abstract}


\section{Introduction}

Shared micro-mobility is increasingly emerging as a key component of a sustainable urban transportation system, providing a flexible and low-emission alternative to private vehicles. By optimizing and adopting first- and last-mile connectivity using services such as bicycles and e-bike sharing systems. However, its effectiveness depends on when, where, and under what environmental conditions people choose to use it. Considering the large-scale adoption of these services \parencite{hosseinzadeh2021micromobility}, detailed research is needed to evaluate the temporal and environmental factors that influence the use of shared micro-mobility services and account for difference in origin-destination (OD) locations.

The development and evolution of shared micromobility systems have focused substantially on improving the supply of shared micromobility systems through station placement, infrastructure investment, fleet expansion, and operational optimization \parencite{forma20153,schuijbroek2017inventory}. These approaches are essential to advance the allocation of services and system efficiency. However, the presence of infrastructure alone does not ensure the adoption of these services. The success of shared bikes depends equally on the demand side, implying the choice of people to incorporate these services into their daily travel routine \parencite{maas2022motivators, eccarius2020adoption}.  Therefore, understanding the demand side perspective has evolved recently in several academic disciplines studying the factors influencing these services, including usage patterns \parencite{bai2020escooter, shah2023escooter}, intentions and behavioral change \parencite{eccarius2020adoption,kaplan2015bikesharing, dadashzadeh2026micromobility}, temporal and spatial \parencite{shah2023escooter, mckenzie2020urban}, to name a few. 

Among the different determinants of shared micro-mobility demand, environmental conditions hold an important criterion for direct exposure of users to the atmospheric conditions. Weather conditions such as rain and temperatures have been studied to affect regular commuter riding \parencite{nankervis1999effect}. With increasing climatic changes and weather patterns, understanding the relationships between environmental conditions and mobility behavior is an urgent requirement \parencite{bocker2019weather}. Understanding interrelated dimensions of weather mobility, such as the choice of transport mode, travel distance, duration, and total origin-destination points, is still a gap in the field of weather and mobility. Some studies \parencite{bocker2019weather} and \parencite{liu2015investigating} provides an example of studying some of these factors together in Swedish, Norwegian, and Dutch cities.  

Although shared bikes are promoted as a strategic investment to reduce transport-related emissions and a sustainable alternative, their long-term contribution depends on whether users continue adopting these different modes under changing environmental conditions. The relationship between transport and weather is complex \parencite{chapman2007climate}. Moderate temperature generally motivates attractiveness to cycle, whereas extreme heat can reduce willingness to use the service due to physical constraints \parencite{bean2021does,Zheng2025}. Similarly, rainfall negatively affects ridership due to safety concerns and inconvenience \parencite{bean2021does}.

Beyond meteorological conditions, air quality is another important factor that needs to be studied in influencing public bike sharing \parencite{yoo2024delayed, noland2021scootin,Xu2025,Kuskapan2024}.  The transportation sector contributes to the high level of air pollution \parencite{nieuwenhuijsen2020urban}. Therefore, shifting users from private to shared vehicles is an important strategy to improve the sustainability of the environment. However, as a result of weather conditions, the relationship between air pollution and demand for shared bikes is also complex. Although deterioration of air quality increases the societal need to adopt sustainable alternatives, on the same side, cyclists and e-bike users experience direct exposure to air pollution during travel. \parencite{liang2023examining} have illustrated studies on the effect of air pollution in reducing lung performance, cardiovascular, and respiratory functions. Simultaneously, it reduces visibility, leading to the switch from public bike sharing to enclosed transportation modes, and therefore to an increase in pollutant emission \parencite{fan2021health}. Therefore, the overall dynamics between environmental conditions and shared micro-mobility demand can be termed as avoidance vs. adoption mechanism which is unlikely to be linear. 

Despite growing evidence on the influence of environmental conditions, several important gaps remain. Existing studies exploring the relationship between environmental conditions and shared micro-mobility demand have focused mainly on changes in overall usage levels, such as variations in daily or hourly travel demand in different weather conditions \parencite{bocker2019weather, bean2021does}. This work extends research on the demand-side perspective to explore how environmental conditions shape daily shared micro-mobility flows in Venice. The study analyzes repeated variation across OD pairs for two years and examines nonlinear effects of temperature, rainfall, and PM$_{10}$ concentrations while accounting for vehicle-pass composition and temporal patterns. Here, the unit of analysis is the connection between OD pairs, rather than only the number of trips made within an area. A generalized additive mixed model (GAMM) is used to represent the non-linearity of environmental relationships across seasons, providing a flexible framework for understanding how climatic conditions influence sustainable mobility behavior. 


\section{Study Area}

The study is conducted within the Municipality of Venice, located in the Veneto region of northeastern Italy. It is a unique case study for various reasons. First, the shared micro-mobility demand in Venice emerges from the interaction of highly heterogeneous natural and urban landscapes \parencite{Ramieri2000Venice}. Second, unlike other European cities with homogeneous road networks, the city comprises a historic island city, mainland districts with more than 100 km of bike lanes, and the barrier island of Lido, each supporting fundamentally different behavior for the demand for transportation. The coexistence of contrasting environments within a single metropolitan area creates a substantial spatial variation in both mobility opportunities and travel demand. The territorial peculiarities (land and sea) make it an important model of urban sustainability. 

Shared bikes and electric bikes serve different mobility functions in these urban environments. In mainland districts, they primarily support first- and last-mile access to public transport and traditional commuting, while within Lido they serve as recreational and tourist travel. In order for this to occur, the determinants of shared micromobility demand are unlikely to be spatially uniform. Understanding how environmental conditions influence mobility, therefore, creates a requirement for an analytical framework to account for localized differences in travel behavior.

\begin{figure}[H]
    \centering
    \includegraphics[width=1.0\linewidth]{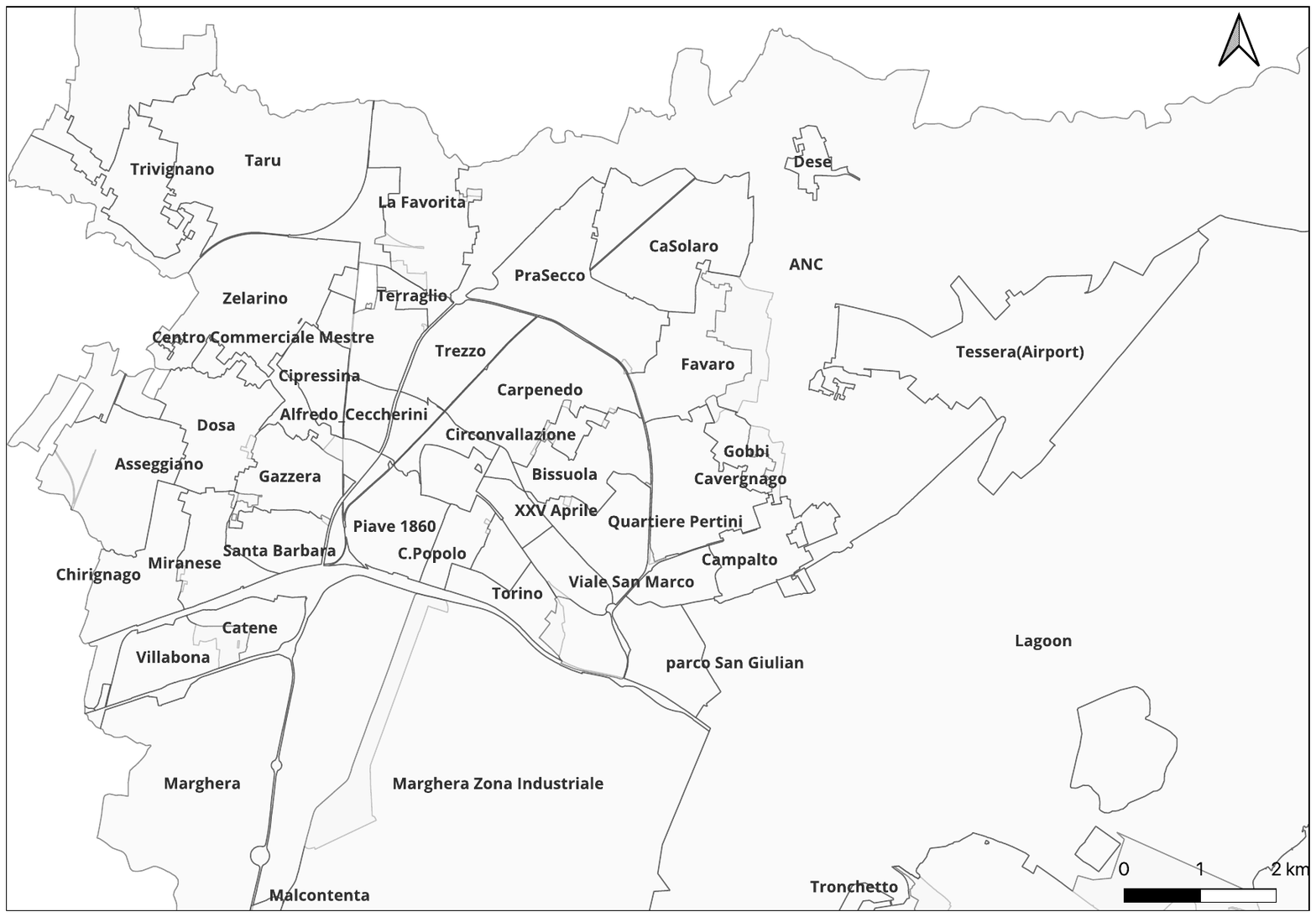}
    \caption{Reconstructed spatial zones in the Municipality of Venice}
    \label{fig:Map}
\end{figure}

To capture these spatial differences, the municipality region was disintegrated into 50 zones specifically designed for the study (Figure ~\ref{fig:Map}). A custom zonal system allows individual trips to be aggregated into daily OD flows while accounting for spatial interactions throughout the municipality. This framework enables a comprehensive integration of environmental conditions, temporal variation, and user composition within a unified modeling approach.  
\section {Methodology}
\subsection{Trip Data Preparation}
The study uses shared micromobility trip records from within the Venice Municipality for the period between January 2024 and December 2025. Each observation includes the start and end coordinates of the trip, time, vehicle type, subscription category, duration of the trip, and travel distance. Instead of independently analyzing individual trips, the objective of this study is to examine how environmental conditions influence the demand for daily mobility in the city. Therefore, the unit defined is the daily OD flow, representing the number of trips between a pair of zones on a given day.  

To describe mobility on a spatial scale for statistical modeling, individual trip origins and destinations are linked to the 50 zones created earlier. Existing administrative divisions within the municipality were either too aggregated to capture local behavior or were excessively detailed for OD modeling \parencite{ComuneVeneziaAreeAgricole2025}. Aggregating trips to daily OD flows reduces the random variability associated with individual travel decisions while retaining the temporal and spatial dynamics required to evaluate environmental influences. Simultaneously, it helps to integrate with daily meteorological data, air-quality measurements, and user composition in a homogeneous approach. The origin and destination coordinates are linked to the 50 spatial locations by creating a shape file for the studied area using the geographic information system (GIS) procedures. Trips outside the defined area or lacking geographic information are excluded from the analysis. 

To account for changes in user composition over time and location, the pass(or subscription) type and vehicle type are combined to create a vehicle-pass category. Categories with few observations were aggregated into the "other" category. Similar products, such as Times Pass and Premium Pass, are grouped into a broader subscription category. The daily proportion of trips associated with each vehicle pass classification is then calculated at the OD day level. 

\subsection{Environmental Data}

To assess the relationship between atmospheric conditions and shared micromobility demand, daily environmental variables are included. Three environmental indicators are used: temperature, rainfall, and particulate matter with an aerodynamic diameter less than 10 μm (PM$_{10}$). These variables represent complementary dimensions of the urban environment that can influence travel behavior. Temperature and precipitation capture short-term weather variability, while PM$_{10}$ reflects daily fluctuations in urban air quality that tend to influence outdoor travel behavior.

The mean daily air temperature for 2024 and 2025 is retrieved from the Open-Meteo historical weather archive. Temperature was analyzed as a continuous variable because previous studies have consistently reported nonlinear relationships with weather conditions and cycling demands, and with extremities of temperature, the ridership decreases.

\begin{figure}[H]
    \centering
    \includegraphics[width=0.82\linewidth]{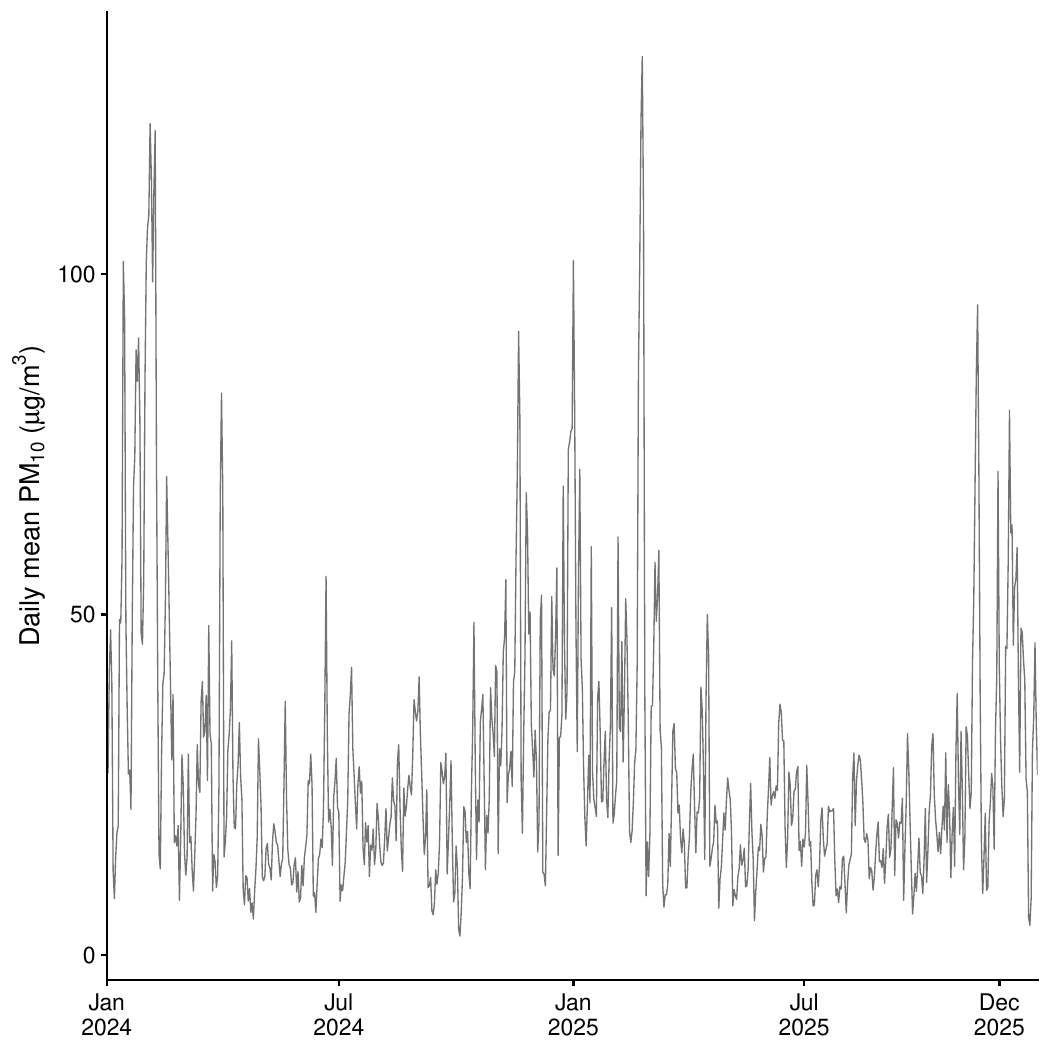}
    \caption{Daily mean PM$_{10}$ concentrations in Venice during the study period (2024–2025)}
    \label{fig:pm10_flow}
\end{figure}

Figure (\ref{fig:pm10_flow}) shows the temporal variation in the mean daily PM$_{10}$ concentrations during the study period. Daily PM concentrations ($\mu g$ / m $^3$) obtained from the Veneto Regional Environmental Protection Agency (ARPAV) include the monitoring of six stations located in the Municipality of Venice: Sacca Fisola, Rio Novo, Malcontenta, Parco Bissuola, Via Beccaria, and Via Tagliamento. Daily measurements are averaged across all stations to obtain a single estimate of ambient air quality for Venice for each day. Reduces the influence of local station-specific variation (also because they are not too far) while considering overall day-to-day variations in pollution levels across the study area.

Rainfall measurements (mm) are classified into four categories: no rain (0 mm), light rain (0.1 -2.5 mm), moderate rain (2.6-10 mm) and heavy rain ($>$10 mm). The approach helps interpret the estimated effects for better comparisons with dry weather. Precipitation data are also obtained from the  (ARPAV), which operates meteorological and air-quality monitoring stations in Veneto.
The environmental data are linked to all the OD flows that occur on the corresponding date. 

Additional temporal variables, including the type of day (weekday/weekend), calendar year, and day of year, are derived from the trip data. The exploratory analysis shows a high localized demand in the Lido zones, particularly between L. Elisabetta and L. Sandro Gallo during August and September. To handle this peak, a binary variable (named as Lido peak) is constructed that marks the observations observed during August and September in these regions and zero otherwise.  

\subsection{Analysis Procedure}

The study employs the generalized additive model for negative binomials (GAMM) to examine the association between environmental and behavioral conditions and daily shared micromobility lows in Venice. Daily OD trips are count data with substantial overdispersion and are expected to vary non-linearly with environmental exposure. The GAMM framework combines the flexibility of a nonparametric smooth function with random effects and a count-data distribution, implemented in the mgcv R package (\parencite{wood2015package}).
The outcome variable is the daily number of trips observed for each OD pair, which follows a negative binomial distribution. Smooth functions are used to model temperature, PM$_{10}$ concentration, and seasonal variation using a day-of-year cyclic smooth function to maintain continuity at the beginning and end of the annual cycle.

\begin{equation}
\log(\mu_{ijt})
=
\beta_0
+
\boldsymbol{\beta}^{T}\mathbf{X}_{ijt}
+
\sum_{m=1}^{M} f_m(Z_{m,ijt})
+
b_{ij},
\end{equation}

\[
b_{ij} \stackrel{\text{i.i.d.}}{\sim}
\mathcal{N}(0,\sigma_b^2)
\]

where $\mu_{ijt}$ denotes the expected daily flow between the origin zone $i$ and the destination zone $j$ on day $t$. The vector $\mathbf{X}_{ijt}$ denotes the parametric covariates, including vehicle-pass categories, rainfall category, weekend, calendar year, and the Lido peak indicator.
The variables \((Z_{m,ijt})\) represent continuous predictors modeled using smooth functions, namely temperature, PM$_{10}$ concentration, and seasonal variation using a day-of-year cyclic spline to maintain continuity at the beginning and end of the annual cycle. In contrast, rainfall is modelled as a categorical variable (no rain, light, moderate and heavy) rather than s a smooth term. This decision was based on the highly skewed nature of the rainfall data (median at 0mm) with most days as dry ($63.3\%$) and few observations under moderate and heavy rainfall. Similar categorical representations of rainfall have been used in transport studies studying weather effects on travel demand \parencite{li2026exploring, kimpton2022weather}. The functions $(f_m(\cdot))$ denote the corresponding smooth effects estimated from the data. The term $b_{ij}$ denotes the random effect associated with each origin--destination pair. It is assumed to be independent and identically distributed Gaussian random variables with mean zero and variance $\sigma_b^2$, allowing each OD-pair to have its own baseline level of daily demand. The random effect variance is estimated using penalized likelihood, resulting in partial pooling of OD-pair effects towards the overall mean. 
The response variable is assumed to follow a negative binomial distribution with a logarithmic link function. In the case of vehicle pass category for better interpretability, the model retains most observed categories, namely, bike pay-as-you-go (PAYG), e-bike PAYG, e-bike subscription and bike monthly. A binary adjustment variable ("Lido Peak") is included to account for the recurring increase in mobility demand observed during August and September between L.Elisabetta and L.Sandro Gallo (Figure \ref{fig:fitted_noLIDO}) location-specific term accounts for the localised variation in demand that is not completely explained by environmental, temporal and behavioural covariates included in the model. 

\section{Results}

\subsection {Descriptive outcomes}
The data set comprises 503,008 individual trips recorded between January 2024 and December 2025. These trips are aggregated by date, origin zone and destination zone, resulting in 158,401 OD-day observations across 1,679 unique pairs of OD over 731 days. Daily trips vary substantially between zones, ranging from 1 to 691. The mean travel distance observed over 2 years was 2.24 km, while the duration was 12.40 minutes. The data suggests that shared micromobility services are used primarily for first- and last-mile connectivity. 

Environmental factors vary throughout the study. Table ~\ref{des} shows that the annual mean daily temperature observed is 15.30 \degree C, ranging from -0.4\degree C to 30.5\degree C. Daily rainfall averaged 2.06 mm, with a median of 0 mm, indicating mostly dry days, although a few observations of extreme precipitation exceeded 130 mm. The mean concentration of daily particulate matter (PM) 10 is also observed as an indicator that varies from 2.67 \(\mu g/m^3\) to 132 \(\mu g/m^3\), while the threshold is 50 \(\mu g/m^3\), as mentioned on the ARPAV website.

\begin{table}[htbp]
\centering

\label{tab:descriptive}
\begin{tabular}{lccccc}
\hline
Variable & Mean & SD & Median & Min & Max \\
\hline
Daily OD flow          & 3.16  & 9.09  & 1.00  & 1.00  & 691.00 \\
Trip distance (km)     & 2.24  & 1.61  & 1.89  & 0.00  & 13.6 \\
Trip duration (min)    & 12.40 & 13.80 & 9.22  & 0.00  & 240.00 \\
Temperature ($^\circ$C)& 15.30 & 7.66  & 15.50 & -0.40 & 30.50 \\
Rainfall (mm)          & 2.06  & 7.47  & 0.00  & 0.00  & 132.00 \\
PM$_{10}$ \((\mu g/m^3)\)       & 26.8  & 19.3  & 21.5  & 2.67  & 132.00 \\
\hline

\end{tabular}
\caption{Descriptive characteristics of the origin--destination daily flow dataset (2024--2025).}
\label{des}
\end{table}

Together, the descriptive outcomes show substantial variability in both ambient and mobility demand throughout the study period. The variability indices the use of flexible nonlinear modeling approach to examine the changes in OD flows associated with weather and air quality.

The use of e-bikes dominates the use of vehicle passes (Table 2). The pay-as-you-go (PAYG) e-bike users represented the highest share of trips (36.1\%), followed by users of e-bike subscription (27.3\%), which consists of time and premium pass. Bike month-card / daily-pass users represented 22.6\% of trips, while PAYG and bike subscription users represented 8.4\% and 5.2\%, respectively. In general, 60\% of trips were recorded on electric bikes, indicating a strong preference for shared mobility assisted by electricity within the study area. The predominance of electric vehicles is one of the most preferred modes within Venice throughout the study. Consequently, changes in vehicle-pass categories can influence overall demand independently of environmental conditions, motivating their inclusion as adjustment variables in the regression model.
\begin{table}[H]
\centering

\label{tab:VP_category}

\begin{tabular}{lcc}
\hline
Vehicle-pass category & Trips (n) & Trips (\%) \\
\hline
E-bike PAYG                  & 181,567 & 36.1 \\
E-bike Subscription          & 137,086 & 27.3 \\
Bike Month Card/Daily Pass   & 113,731 & 22.6 \\
Bike PAYG                    & 42,337  & 8.4 \\
Bike Subscription            & 25,952  & 5.2 \\
Other                        & 2,335   & 0.5 \\
\hline
Total                        & 503,008 & 100.0 \\
\hline
\end{tabular}
\caption{Distribution of trips by vehicle-pass category.}

\end{table}

In modeling the daily counts, vehicle-pass categories are expressed as proportions of trips within each OD-day observation rather than absolute counts. This approach restricts the vehicle-pass effects from confounding with the total number of trips, and allows the model to capture differences in user composition independently of overall demand.

\begin{figure}[H]
    \centering
    \includegraphics[width=0.9\linewidth]{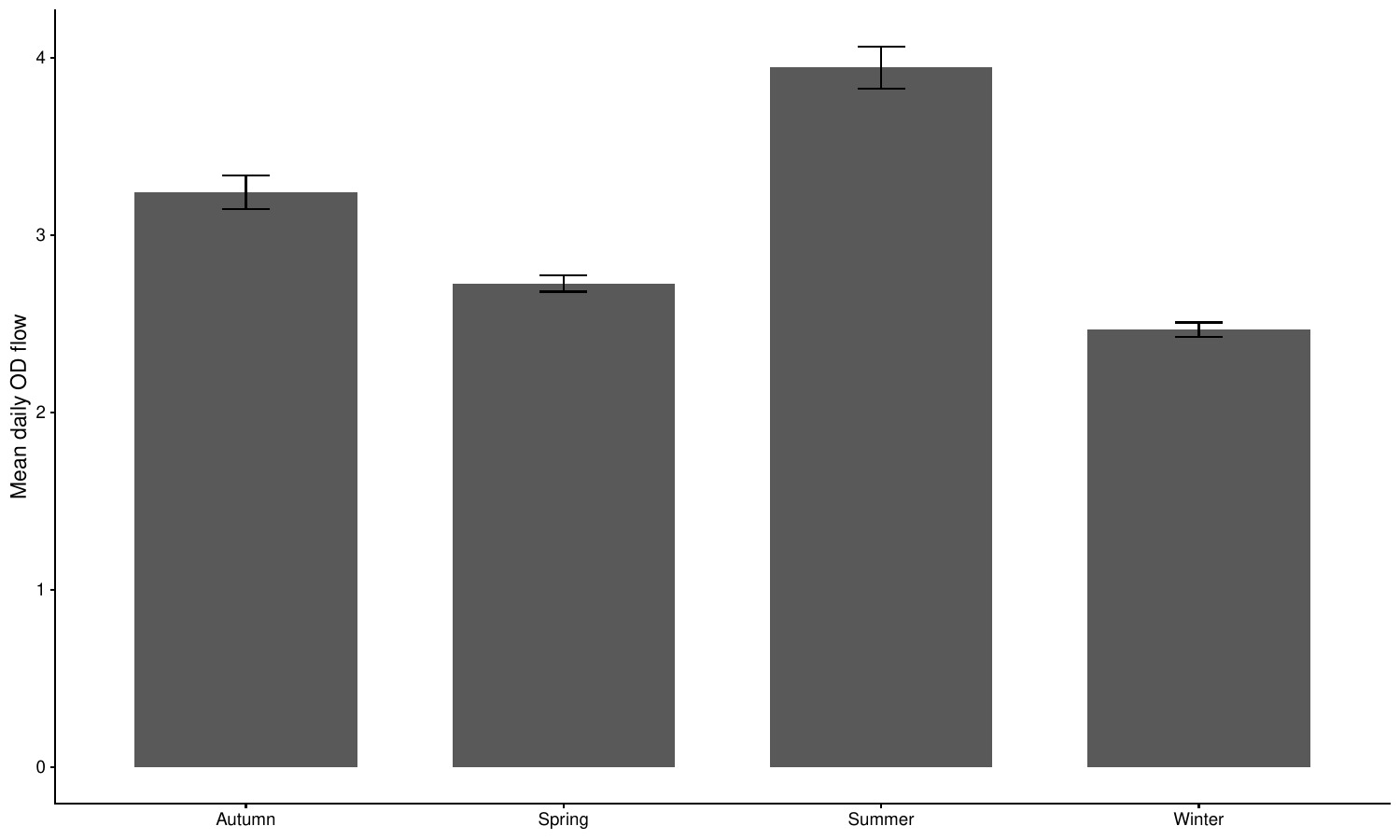}
    \caption{Mean of seasonal variation in the daily demand}
    \label{fig:seasonal}
\end{figure}

The seasonal variation in mobility demand depicts a noticeable pattern (Figure ~\ref{fig:seasonal}), with the highest mean OD flows observed during summer (3.95 trips per OD pair) and the lowest during winter (2.47 trips per OD pair). These findings suggest that the shared demand for micromobility is substantially influenced by seasonal conditions, with warmer months associated with increased travel activity.

The descriptive seasonal pattern further shows that demand is unlikely to remain constant throughout the year. The seasonal effects shown in the descriptive analysis are not included in the model. Instead, seasonality is represented as a cyclic smooth function of day-of-year and through season-specific temperature smooths, allowing the seasonal effect to vary continuously throughout the year. This approach avoids redundancy between highly correlated temporal variables. 

\begin{figure}[H]
    \centering
    \includegraphics[width=1.2\linewidth]{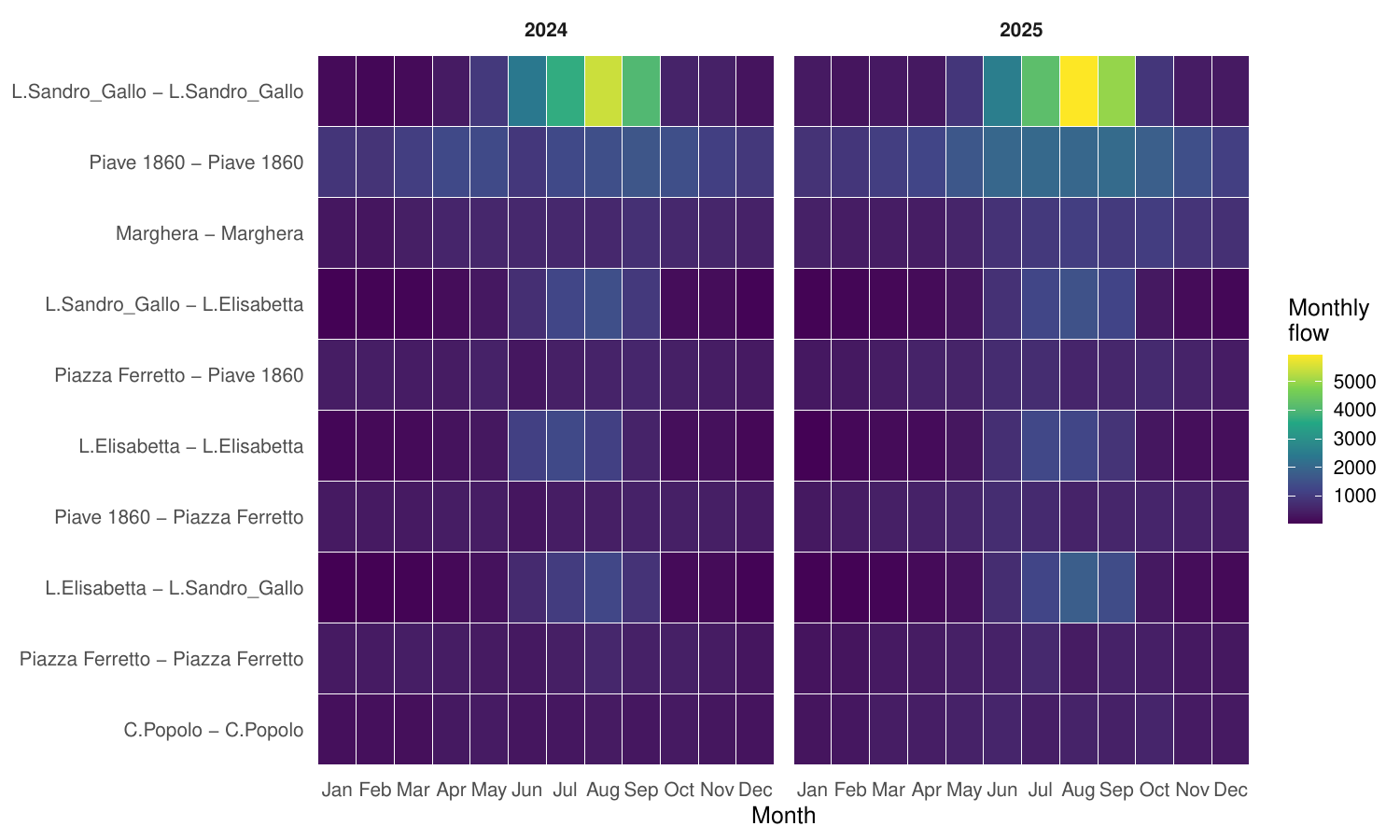}
    \caption{Monthly mobility demand for the top OD pairs}
    \label{fig:heatmap}
\end{figure}

Several high-demand zones, such as Piave 1860, Marghera, and Lido, represent seasonal variations. The persistence of the seasonal pattern over two years suggests that the observed increase is unlikely to be random, but instead reflects a high location-specific demand associated with tourism and recreational activity. The zones within the Lido territory show an increase in specific seasons. The monthly flows between L.Elisabetta and L.Sandro Gallo peak during August and September, as shown in Figure ~\ref{fig:heatmap}. The flows between Piave 1860 are also high, but a sudden change in August and September is observed only for L. Elisabetta and L.Sandro. This yearly pattern was not observed in any other origin or destination location and motivated the need for a specific Lido peak binary variable. 

The descriptive analyses show an independent substantial variation in the shared demand for micromobility across environmental conditions, vehicle-pass categories, and specific pairs of OD. To jointly evaluate the association among them, several specifications of the generalized additive mixed model (GAMM) with a negative binomial distribution were fitted (Appendix A). 

\subsection{Generalized additive mixed model results}

The final GAMM explains 81.6\% of the deviation in the daily OD flows, providing a good fit for the data. The high proportion of deviance explained obtained indicates the selected environmental, behavioral and temporal variables to capture most of the variability in daily shared micro-mobility demand. The inclusion of the Lido peak indicator improved the fit of the model. 
This confirms the location as a prime tourist spot in specific months of the year. After adjustment for environmental conditions, temporal variation, vehicle-pass composition, and OD pair heterogeneity, the Lido peak corridor remained strongly associated with increased mobility demand. This corridor comprised two OD pairs, L. Sandro Gallo - L.Sandro Gallo and L.Elisabetta - L.Sandro Gallo. The daily flows during August and September in those zones are more than three times higher than expected relative to other OD pairs. These results indicate that the variations of the user composition contribute to the difference in mobility demand beyond environmental conditions. Specifically, the consistent positive association observed for e-bike users suggests that electrical vehicles are potentially less constrained by physical conditions, therefore supporting daily usage under changing outer conditions. 

The composition of the Vehicle-pass is associated with the intensity of the flow. The increase in demand is mainly driven by the users of e-bikes relative to the normal users of bikes. OD pairs with a higher proportion of PAYG e-bike users exhibited significantly higher flows by 8\% (RR = 1.08, 95\% CI: 1.06--1.11), while higher proportions of e-bike subscription users and bike month-card users were associated with moderate increases by 5\% (RR = 1.05, 95\% CI: 1.03--1.08) demand. In contrast, the proportion of PAYG bike users was not significantly associated with daily flow levels. 

Temporal and weather-related factors are also significantly associated with daily mobility demand. After adjustment for other covariates, weekend flows are approximately 8\% lower than weekday flows (RR = 0.92, 95\% CI: 0.91--0.92). In contrast, daily OD flows in 2025 were 15\% higher than those observed in 2024 (RR = 1.15, 95\% CI: 1.14--1.16), indicating a continued increase in shared micro-mobility usage during the study period.

\begin{table}[H]
\centering
\caption{Parametric effects from the final generalized additive mixed model of daily OD flows.}
\label{tab:gamm_results}

\small

\begin{tabular}{p{6cm}ccc}
\hline
Variable & Rate Ratio & 95\% CI & p-value \\
\hline
Lido peak (August--September) & 3.27 & (3.10, 3.45) & \textless 0.001 \\
E-bike PAYG proportion & 1.08 & (1.06, 1.11) & \textless 0.001 \\
Bike PAYG proportion & 0.99 & (0.97, 1.03) & 0.854 \\
E-bike subscription proportion & 1.05 & (1.03, 1.08) & \textless 0.001 \\
Bike month-card proportion & 1.05 & (1.03, 1.08) & \textless 0.001 \\
Weekday (R) &  &  &  \\
Weekend & 0.92 & (0.91, 0.92) & \textless 0.001 \\
Year 2024 (R) &  &  &  \\
Year 2025 & 1.15 & (1.14, 1.16) & \textless 0.001 \\
No rain (R) &  &  &  \\
Light rain & 0.96 & (0.95, 0.97) & \textless 0.001 \\
Moderate rain & 0.88 & (0.87, 0.90) & \textless 0.001 \\
Heavy rain & 0.78 & (0.76, 0.79) & \textless 0.001 \\
\hline
\end{tabular}

\vspace{0.2cm}

\begin{minipage}{0.95\textwidth}
\footnotesize
Rate ratios were obtained by exponentiating model coefficients. The final model explained 81.6\% of the deviance in daily OD flows. Temperature, PM$_{10}$ and day-of-year were modelled using smooth functions and are presented separately in Figures X--Y.
\end{minipage}

\end{table}

Rainfall is associated with a continuous reduction in mobility demand. Compared to dry days, light rainfall was associated with a slower decrease in flow (RR = 0.96, 95\% CI 0.95--0.97), while moderate and heavy rainfall resulted in relatively greater reductions (RR = 0.88, 95\% CI 0.87--0.90 and RR = 0.78, 95\% CI 0.76--0.79, respectively). The consistent decline in flow with increasing rainfall intensity suggests that adverse weather conditions are an important factor in shared micro-mobility use.

\begin{table}[htbp]
\centering
\caption{Approximate significance of smooth terms in the final GAMM.}
\label{tab:smooth_terms}

\small

\begin{tabular}{lccc}
\hline
Smooth term & EDF & F-statistic & p-value \\
\hline

Temperature (Autumn) & 7.61 & 25.79 & $<0.001$ \\
Temperature (Spring) & 5.69 & 11.04 & $<0.001$ \\
Temperature (Summer) & 3.75 & 5.31 & $<0.001$ \\
Temperature (Winter) & 3.74 & 6.98 & $<0.001$ \\

PM$_{10}$ (Autumn) & 5.99 & 5.90 & $<0.001$ \\
PM$_{10}$ (Spring) & 4.55 & 11.61 & $<0.001$ \\
PM$_{10}$ (Summer) & 3.30 & 1.41 & 0.136 \\
PM$_{10}$ (Winter) & 5.99 & 11.93 & $<0.001$ \\

Day of year & 17.28 & 45.25 & $<0.001$ \\

OD-pair random effect & 1143.94 & 171.04 & $<0.001$ \\

\hline
\end{tabular}

\vspace{0.2cm}

\begin{minipage}{0.95\textwidth}
\footnotesize
EDF = estimated degrees of freedom. Larger EDF values indicate increasingly non-linear relationships between the predictor and daily OD flow.
\end{minipage}

\end{table}

The smooth temperature effect shows statistical significance in all four seasons, indicating the non-linear association between temperature and daily mobility demand throughout the year. The estimated degrees of freedom ranged from 3.7 to 7.6, suggesting increasingly complex relationships during fall and spring. Similarly, the season-specific smooth functions for PM$_{10}$ are significant during autumn, spring, and winter, but not during summer. The cyclic spline effect of day-of-year is also significant, confirming the presence of seasonal variation beyond that captured by the observed meteorological variables.

The random effect of OD pairs has a substantially large EDF (1143.94), reflecting the high variation in baseline demand between OD pairs. The penalized random-effects specification applied partial pooling, shrinkage estimates for sparsely observed OD pairs towards the overall mean while allowing pairs with more observation retain more more distinct baseline effects. 

\begin{figure}[H]
    \centering
    \includegraphics[width=1.0\linewidth]{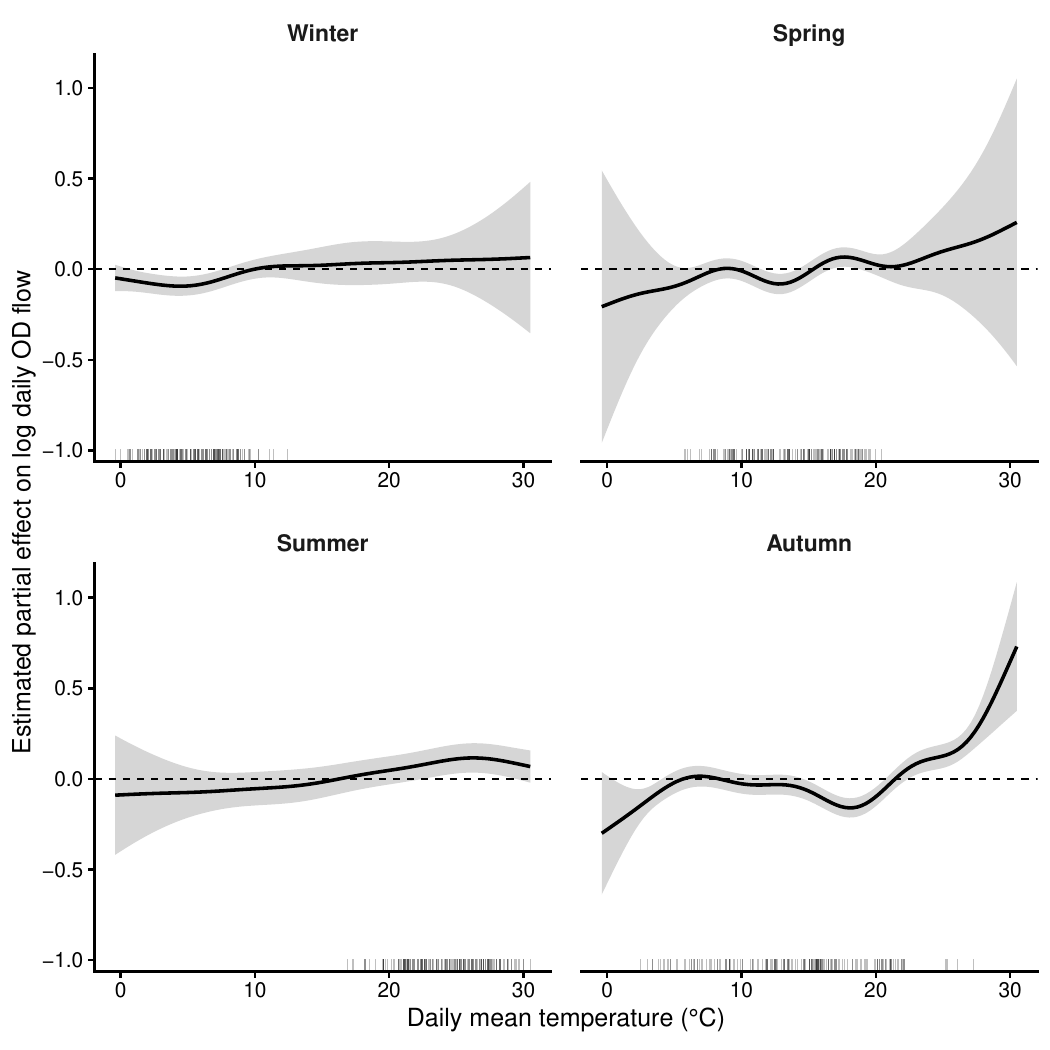}
    \caption{Estimated season-specific smooth effects of daily mean temperature on daily OD flow obtained from GAMM}
    \label{fig:placeholder}
\end{figure}

Figure 5 shows different nonlinear relationships between temperature and daily OD flows across seasons. During winter, the estimated effect remains relatively stable throughout the observed cold temperature range. In contrast, spring and fall exhibit stronger positive associations as temperature increases, while summer illustrates a more gradual increase. The wider confidence interval in gray shades at the lower and upper ends of the temperature range reflects the smaller number of observations under extreme weather conditions. 

\begin{figure}
    \centering
    \includegraphics[width=1.0\linewidth]{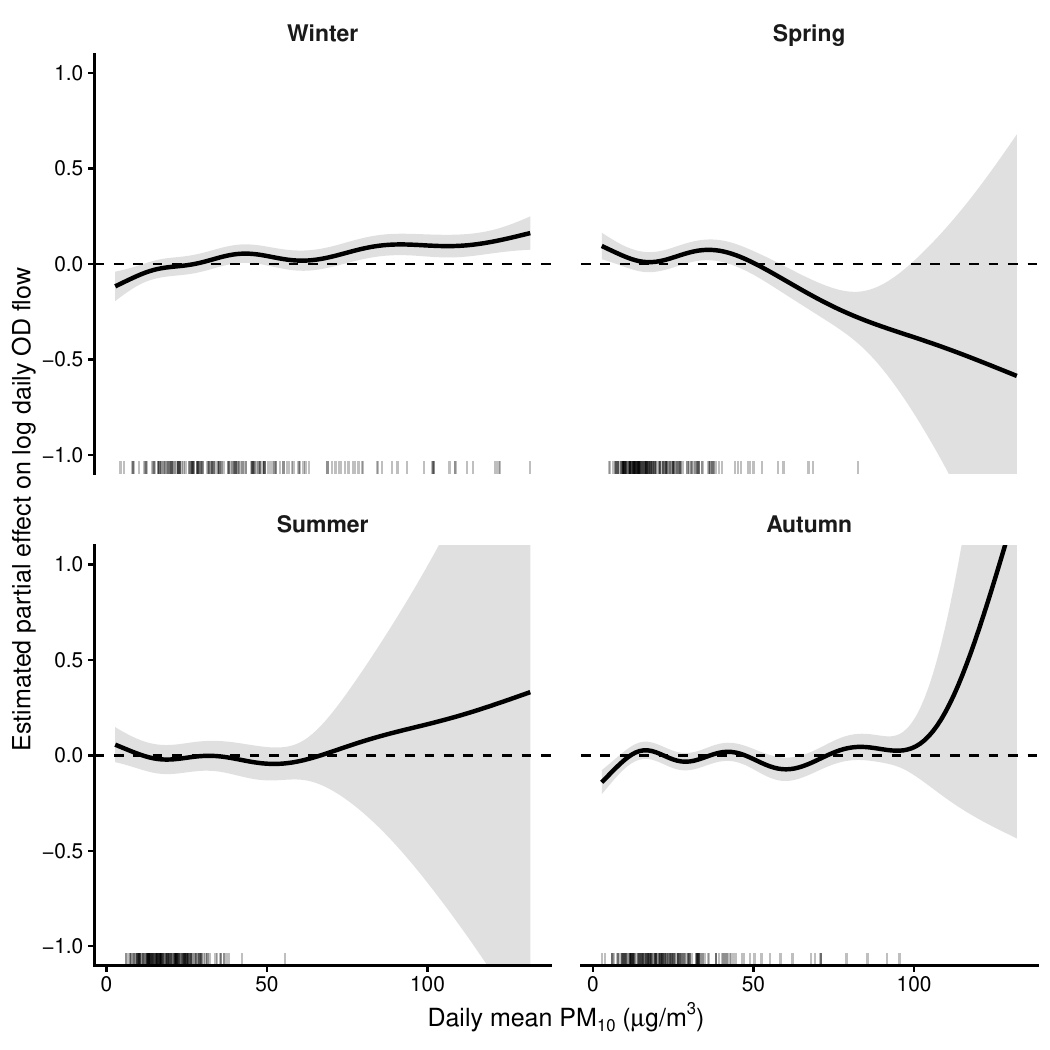}
    \caption{Season-specifc PM$_{10}$ smooth results from GAMM}
    \label{fig:pm10_season}
\end{figure}

Figure (\ref{fig:pm10_season}) indicates that the relationship between PM$_{10}$ and shared micromobility demand does not exhibit a common pattern between seasons.  Spring shows a gradual decline in demand with an increasing mean of PM$_{10}$, while winter remains relatively stable and summer exhibits weak evidence of association. The sharp increase in autumn occurs during high concentrations of PM$_{10}$ with greater uncertainty in the smooth estimated due to relatively few observations at these high levels. These findings are therefore consistent with the discussion that air pollution creates an avoidance vs. adoption mechanism on shared mobility use rather than a monotonic association. 

\section{Conclusion and Discussion}

This study examined how environmental conditions influence shared micro-mobility demand from an OD flow perspective. The finding reveals that weather and air quality conditions are significantly associated with daily mobility flows over two years of Venice trip records. They highlight the importance of environmental sensitivity in the domain of shared micro-mobility studies. This work illustrates that the adoption of shared bikes and electric bikes depends not only on service availability, but also on the usage pattern that is affected by external conditions.

Temperature shows a nonlinear association with demand between seasons, which is consistent with previous research that the effect of weather on cycling behavior is complex and context driven \parencite{bean2021does, bocker2019weather}. Users travel in a manner perceived based on comfort and environmental needs, which is consistent with the concept of weather conditions that encourage cycling and limit practice in extreme temperatures to physical discomfort and move to enclosed vehicles \parencite{Zheng2025}. 

Rainfall showed a consistent negative effect with shared micro-mobility demand, with stronger effects during heavy and moderate rain. The results are similar to those of previous studies that reported precipitation as the strongest environmental factor affecting the use of the service due to comfort, visibility, and safety problems \parencite{bean2021does, noland2021scootin}. Unlike rainfall, air quality contributes as a more complex relationship with demand for shared bikes. Significant nonlinear associations are observed during autumn, winter, and spring, but not in summers, indicating that the response to pollution emerges as a competing behavior, which is that while sustainable transport is promoted in response to pollutants, individuals tend to avoid exposure and shift modes of travel \parencite{Xu2025, liang2023examining}.

Beyond environmental factors, the main contribution of the study is focus on the OD-flow perspective. Modeling mobility at the OD level allows one to take into account localized variations in trip flows, which might not be the case in aggregate trip counts. The specific heterogeneity observed across OD pairs suggests that shared micro-mobility demand is strongly influenced by the characteristics of specific routes and locations. We observed this in the Lido regions for the study, specifically in Elisabetta and Sandro Gallo where the flows increased substantially during August and September. After several specifications in the model on environmental and temporal patterns, this corridor reflected a unique seasonal mobility demand. It highlights the importance of inclusion of local factors such as, tourism and recreational travel in shared micro-mobility studies. A continuous growth from 2024 to 2025 is observed in adoption of the service. Apart from the factors consisdered, we believe that the overall increase is also related to familiarity with shared mobility services, changes in travel behavior, or acceptance of these services as part of daily transportation. The results also show a large number of rides made on electronic bikes. This indicates that e-bikes can reduce barriers related to comfort and travel distance, potentially increasing adoption under less favorable conditions. 
\printbibliography
\appendix

\section{Model specification and comparison}

Several generalized additive mixed-model (GAMM) specifications were evaluated during model development. The models differed in the representation of PM$_{10}$ and the inclusion of the Lido peak adjustment variable. All models included vehicle-pass composition, rainfall category, weekend, calendar year, season-specific temperature smooths, a cyclic smooth of day-of-year, and an OD-pair random effect. Model selection was based on Akaike's Information Criterion (AIC), effective degrees of freedom (EDF), adjusted $R^2$, deviance explained, diagnostic plots, and model interpretability.

\begin{table}[H]
	\centering
	\caption{Comparison of the GAMM specifications in fitting model.}
	\label{tab:model_comparison}
	
	\small
	
	\begin{tabular}{p{0.8cm}p{2.5cm}p{1.0cm}rrrrr}
		\toprule
		Model &
		PM$_{10}$ smooth &
		Lido peak &
		EDF &
		AIC &
		\makecell{Deviance\\explained (\%)} &
		Adjusted $R^2$ \\
		\midrule
		
		M1 &
		Overall &
		No &
		1200 &
		532600 &
		81.2 &
		0.455 \\
		
		M2 &
		Season-specific &
		No &
		1213 &
		532490 &
		81.2 &
		0.455 \\
		
		M3 &
		Season-specific &
		Yes &
		1213 &
		530438 &
		81.6 &
		0.464 \\
		
		\bottomrule
	\end{tabular}
	
\end{table}

Table \ref{tab:model_comparison} shows the comparison between models with overall PM$_{10}$ smooth (M1),  PM$_{10}$ association to vary by season (M2) and then with incorporating the Lido peak adjustment variable (M3). From M1 to M2 the AIC reduced by approximately 110 points while leaving the overall explanatory variables same.  M3 shows a substantially larger improvement in model fit by reduction of 2,052 points AIC and increasing the deviance explained from 81.2\% to 81.6\%. Consequently, M3 was selected as the final model used for inference. Across all specifications, the estimated parametric effects remained stable. 

\section{Model diagnostics}
The generalized additive mixed model using the fast restricted maximum likelihood (fREML) estimation procedure implemented in the \textit{mgcv} package. The optimization was terminated with negligible gradients and a full-rank Hessian matrix, indicating numerically stable parameter estimation.

\begin{figure}[H]
    \centering
    \includegraphics[width=1.0\linewidth]{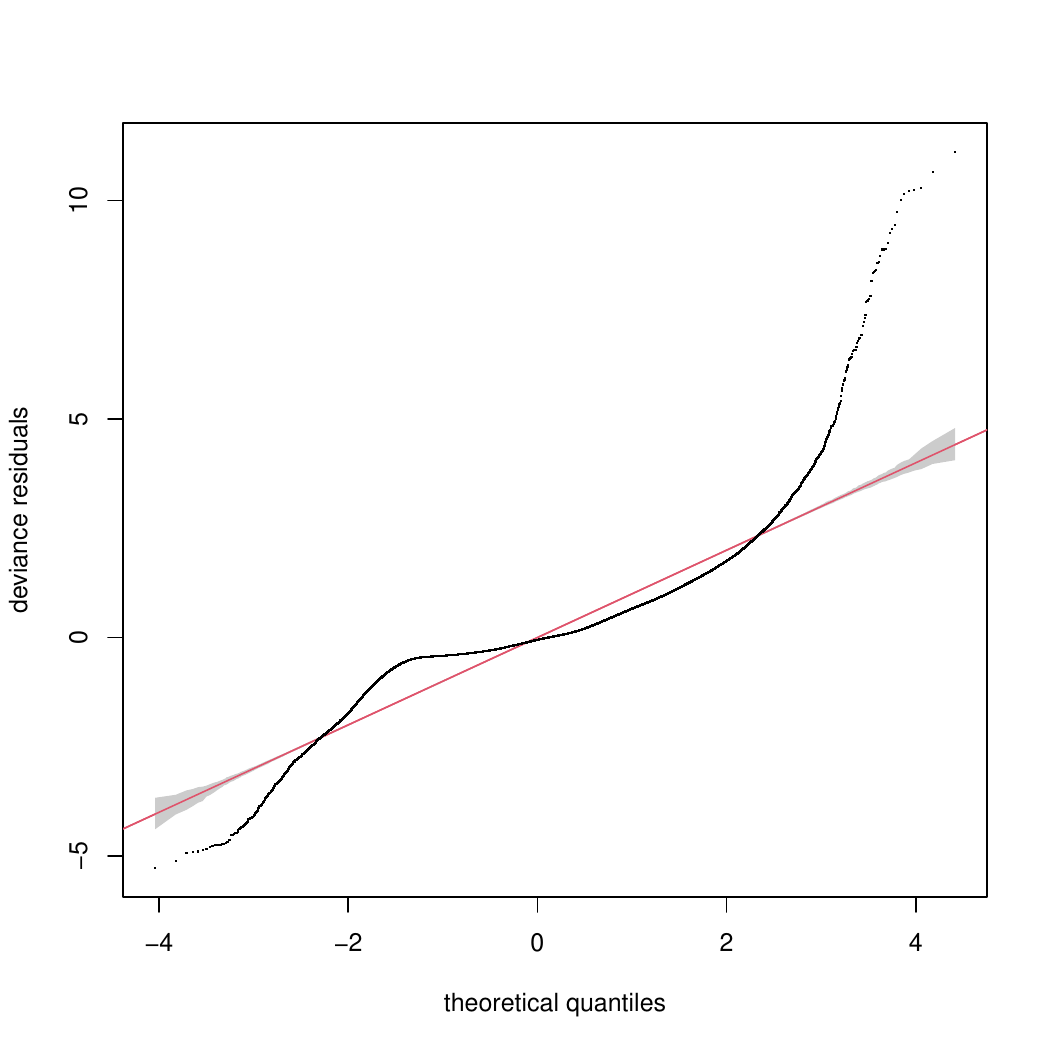}
    \caption{Quantile-Quantile plot of deviance residuals}
    \label{fig:QQplot}
\end{figure}

Residual diagnostics were examined using deviance residuals obtained from the fitted negative binomial GAMM. 
Figure \ref{fig:QQplot} presents a quantile-quantile plot of the deviation residuals. The residuals show a moderate deviation primarily in the upper tails, although they show a distribution well over most of the range. Such deviations are expected in large over-dispersed count datasets and do not necessarily indicate model misspecification. \parencite{augustin2012quantile}. 

\begin{figure}[H]
    \centering
    \includegraphics[width=1.0\linewidth]{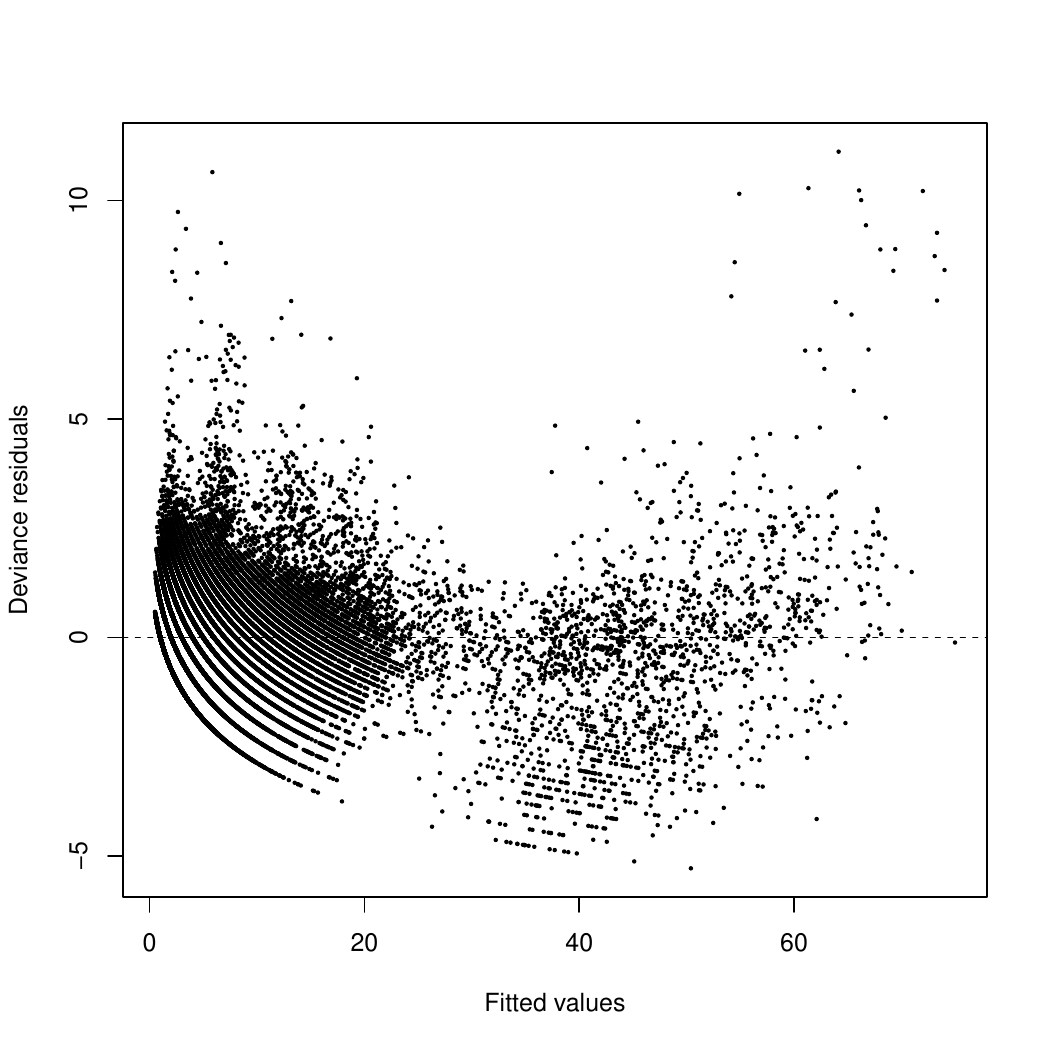}
    \caption{Deviance residuals plotted against fitted values}
    \label{fig:fitted_GAM}
\end{figure}

The relationship between fitted values and the deviation residuals (Figure \ref{fig:fitted_GAM}) shows residuals concentrated around zero throughout the range of fitted values. The plot shows a discrete nature of the count data. The slightly greater dispersion at large fitted values is consistent with the negative binomial distribution used to model daily OD flows. 

\section{Basis-dimension diagnostics}

Table \ref{tab:F1} shows the basis-dimension diagnostics obtained using \texttt{gam.check()}. The estimated effective degrees of freedom (EDF) are below the corresponding basis dimensions (\textit{k'}) for all smooth terms. The k-index values ranged from 0.940 to 0.968, which are almost close to the recommended value of one. Several smooth terms produced relatively small diagnostic p-values. Taking into account the residual diagnostics shown in Appendix E, there was no serious model inadequacy.

\begin{table}[H]
\centering
\caption{Basis-dimension diagnostics obtained using \texttt{k.check()} for the final negative binomial generalized additive mixed model.}
\label{tab:F1}

\begin{tabular}{lcc}
\hline
Smooth term & $k'$ & k-index \\
\hline
Temperature (Autumn)      & 9  & 0.959 \\
Temperature (Spring)      & 9  & 0.959 \\
Temperature (Summer)      & 9  & 0.959 \\
Temperature (Winter)      & 9  & 0.959 \\

PM$_{10}$ (Autumn)        & 7  & 0.968 \\
PM$_{10}$ (Spring)        & 7  & 0.968 \\
PM$_{10}$ (Summer)        & 7  & 0.968 \\
PM$_{10}$ (Winter)        & 7  & 0.968 \\

Day of year               & 18 & 0.940 \\

OD-pair random effect     & 1679 & -- \\
\hline

\end{tabular}
\end{table}

\section{Identification of the Lido peak effect}

The baseline GAMM (without the Lido adjustment variable) explained $81.2\%$ of the deviance in daily OD flows. Following ihe inclusion of Lido variable, the AIC decreased from 532,484.7 to 530,433.7, while the deviance explained increased from $81.2\%$ to $81.6\%$. The reduction in AIC shows an improved balance between model fit and complexity, whereas a slight increase in deviance explained suggests that the additional variable primarily accounts for a localized and recurring demand pattern between L. Elisabetta and L. Sandro Gallo during August and September. Therefore, the Lido peak variable is interpreted as a location-specific adjustment term that accounts for residual variation. 
Figure 9 shows the relationship between daily OD flows observed and the fit values obtained from the GAMM before introducing the Lido peak term. Therefore, a binary Lido peak indicator was constructed to justify the location-specific seasonal demand. 

\begin{figure}[H]
	\centering
	\includegraphics[width=\linewidth]{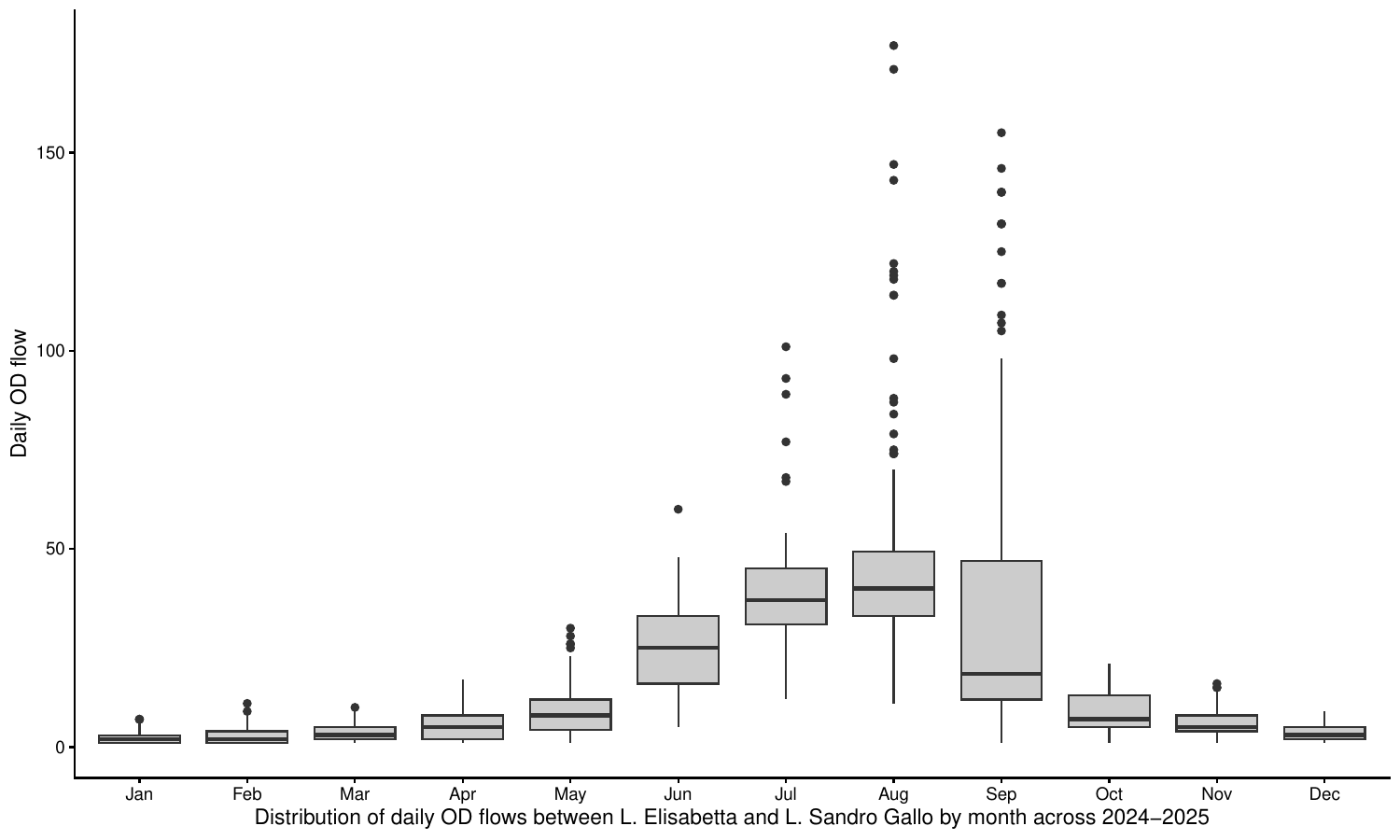}
	\caption{Observed daily OD flows against fitted values from the baseline GAMM (without the Lido adjustment variable).}
	\label{fig:fitted_noLIDO}
\end{figure}

\begin{figure}[H]
    \centering
    \includegraphics[width=1.0\linewidth]{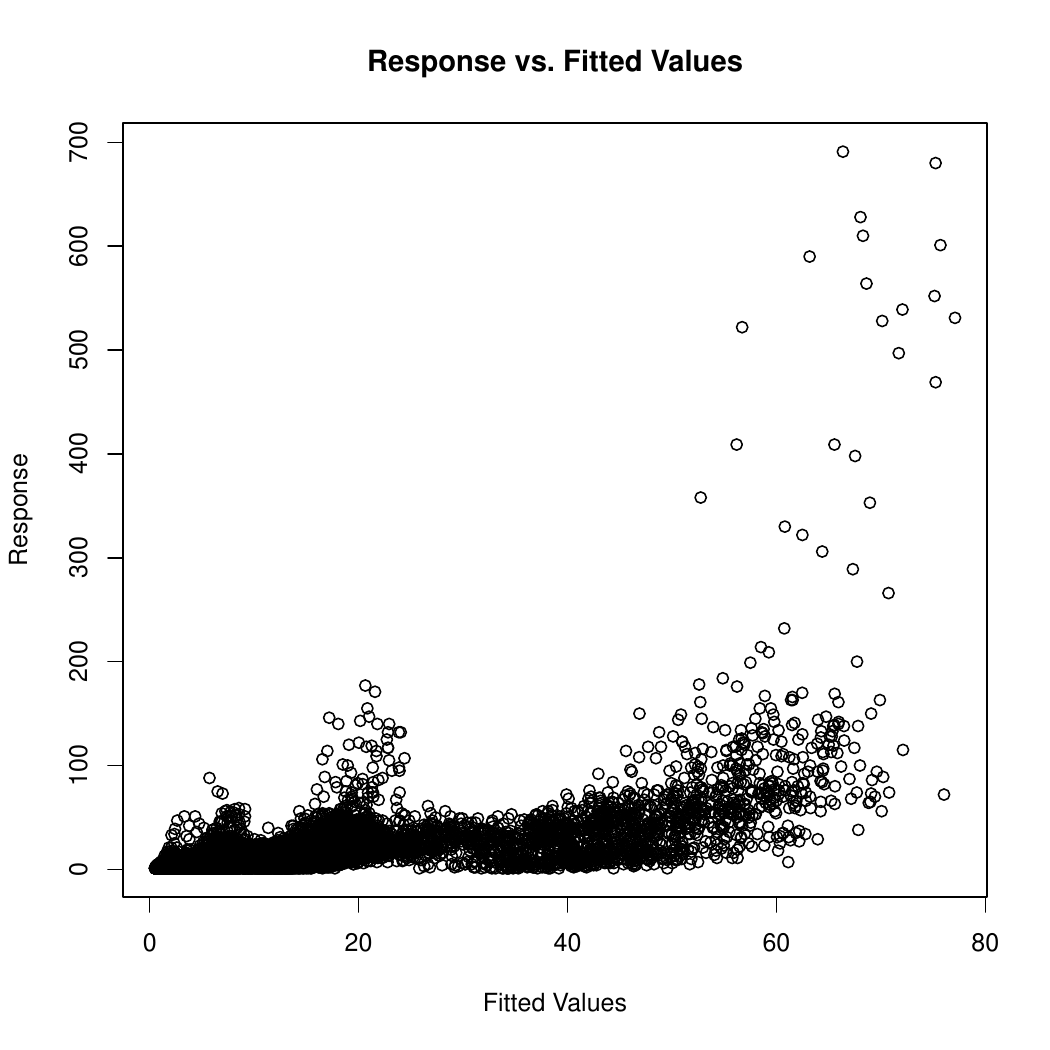}
    \caption{Observed daily OD flows against fitted values from the GAMM before inclusion of Lido peak indicator}
    \label{fig:noLido_GAMM}
\end{figure}

\end{document}